\documentclass[aps,pra,superscriptaddress,12pt,tightenlines,nofootinbib]{revtex4}
\usepackage{amsmath,amsthm,graphicx,amssymb,verbatim,listings}
\usepackage[T1]{fontenc}     
\usepackage{lmodern}         
\usepackage[ pdftex, plainpages = false, pdfpagelabels,
                 pdfpagelayout = useoutlines,
                 bookmarks,
                 bookmarksopen = true,
                 bookmarksnumbered = true,
                 breaklinks = true,
                 linktocpage=all,
                 pagebackref=false,
                 colorlinks = true,
                 linkcolor = BrickRed,
                 urlcolor  = blue,
                 citecolor = BrickRed,
                 anchorcolor = green,
                 hyperindex = true,
                 hyperfigures
                 ]{hyperref}
\usepackage[usenames, dvipsnames, svgnames, table]{xcolor}
\usepackage{xifthen} 

\newcommand{\tr}{\hbox{tr}}

\newcommand{\ket}[1]{{\ensuremath{\left| #1 \right\rangle}}}
\newcommand{\bra}[1]{{\ensuremath{\left\langle #1 \right|}}}
\newcommand{\braket}[2]{{\ensuremath{\left\langle #1 \middle| #2
      \right\rangle}}}

\newcommand{\arxiv}[2][]{\ifthenelse{\isempty{#1}}{\href{http://arxiv.org/abs/#2}{{\tt arXiv:\allowbreak{}#2}}} {\href{http://arxiv.org/abs/#2}{{\tt arXiv:\allowbreak{}#2 [#1]}}}}

\newcommand{\ogm}{\guillemotleft~}
\newcommand{\cgm}{~\guillemotright}
\newcommand{\booktitle}{\textsl}
\newcommand{\MFQM}{\textsl{MFQM}}

\begin{document}

\title{Von Neumann Was Not a Quantum Bayesian}

\author{Blake C.\ Stacey}
\affiliation{Department of Physics,
  University of Massachusetts Boston, 100 Morrissey Blvd.,
  Boston, MA 02125, United States}


\begin{abstract}
Wikipedia has claimed for over three years now that John von Neumann
was the ``first quantum Bayesian.''  In context, this reads as stating
that von Neumann inaugurated QBism, the approach to quantum theory
promoted by Fuchs, Mermin and Schack.  This essay explores how such a
claim is, historically speaking, unsupported.
\end{abstract}

\maketitle

\section{Introduction}

The Wikipedia article on Quantum Bayesianism has claimed since April
2012 that John von Neumann was the ``first quantum
Bayesian''~\cite{Wikipedia2012}.  To a reader acquainted with quantum
foundations and the history of quantum theory, this is a strikingly
odd assertion.  This note explains why the claim is incorrect and
explores how it came to be made.

A ``Quantum Bayesian'' is one who interprets the probabilities arising
in quantum physics according to some variety of the Bayesian view of
probabilities.  Given the profusion of schools of thought under the
Bayesian umbrella, it should not be surprising that a variety of ways
to be some kind of Bayesian about quantum theory has also
arisen~\cite{Baez2003, Barnum2010, Bub2015, Caticha2007,
  Coecke2012, GellMannHartle2012, Goyal2008, Leifer2006, Leifer2013,
  Pitowsky2003, PortaMana2007, Rau2009, Srednicki2005,
  Warmuth2009}.  The most radical approach is
\emph{QBism,} which maintains that all quantum states are expressions
of personalist Bayesian probabilities about potential future
experiences~\cite{Barnum2010, FuchsPerimeter, FuchsSchack2011,
  Timpson, MerminPT, RMP, AJP, Mermin14, Fuchs2014, Fuchs2014b,
  Stacey2014, Mermin14b, Mermin-Vienna, Mermin-Bell}.  For this essay,
I will take the writings of Fuchs, Mermin and
Schack~\cite{FuchsPerimeter, FuchsSchack2011, RMP, AJP, Fuchs2014,
  Fuchs2014b, MerminPT, Mermin14, Mermin14b, Mermin-Vienna,
  Mermin-Bell} as the defining statements of what a QBist \emph{is}
and \emph{is not}.  Furthermore, their union and intersection are
indicative of what a QBist might be, might not be and is not obligated
to entertain.

QBism is the primary focus of the Wikipedia article on
Quantum Bayesianism, and the take-away impression is that John von
Neumann was not just a partisan of the Bayesian lifestyle, but also
the first QBist.  This is an untenable claim.

In this essay, we will not be strongly concerned with which
interpretation of quantum mechanics is ``correct,'' or with what it
might \emph{mean} for an interpretation of quantum mechanics to
\emph{be} ``correct.''  Our focus will instead be on who said what and
when.  However, to evaluate the ``von Neumann was the first Quantum
Bayesian'' claim properly, we need to clarify what a ``Quantum
Bayesian'' world view might be, and QBism, in many ways an extreme
among such views, provides a convenient vantage point.  Therefore, we
will establish the basic notions of QBism, and then in following
sections we will turn to the writings of von Neumann.

\section{QBism}
\label{sec:qbism}
QBism is an interpretation of quantum mechanics which takes as
fundamental the ideas of \emph{agent} and \emph{experience.}  A
``quantum measurement'' is, in QBism, an act which an agent performs
on the external world.  A quantum state is an agent's encoding of her
own personal expectations for what she might experience as a result of
carrying out an action.  This holds true for all quantum states, pure
or mixed; a state without an agent is a contradiction in terms.
Furthermore, each experience is a personal event specific to the agent
who evokes it.  Cabello classifies QBism as a kind of ``participatory
realism,'' similar to the thinking of John Wheeler~\cite{Cabello2015}.

Different authors have emphasized different aspects of QBism.  The
discussions by Barnum~\cite{Barnum2010} and Mermin~\cite{MerminPT,
  Mermin14, Mermin14b, Mermin-Vienna, Mermin-Bell}, and the briefer
remarks by Schlosshauer, Claringbold and
\.{Z}ukowski~\cite{Schlosshauer2014, Zukowski2015}, place their focus
on how QBism gives meaning to the current mathematical formalism of
quantum theory.  Fuchs and Schack have also addressed this
aspect~\cite{FuchsSchack2011}, while in addition pushing forward
technical work which aims to reformulate quantum theory and build it
up anew from explicitly QBist postulates~\cite{RMP}.  Due to the
historical subject matter of this essay, the former will be more
relevant here.\footnote{This is not to say that the more technical
  side of QBism is without historical and philosophical interest.  The
  roots of the mathematics involved go back to
  Schwinger~\cite{Schwinger1960} and Weyl~\cite[\S IV.D.14]{Weyl1931},
  indeed to the very transition from the ``old quantum theory'' to the
  new~\cite[pp.\ 2055--56, 2257--58, 2280]{Fuchs2014}.  And, in the
  SIC representation of quantum states and channels, the Born
  rule---usually written like $p(i) = \tr(\rho E_i)$---and unitary
  evolution---typically written like $\rho(t) = U_t \rho(0)
  U_t^\dag$---take \emph{the same form.}  Both are simple affine
  deformations of the Law of Total Probability~\cite{RMP}.  This
  clarifies that both are \emph{synchronic} relations between
  probability ascriptions~\cite{FuchsSchack2011}.  Alice carries a
  probability distribution for an informationally complete
  measurement, which she uses to summarize her expectations.  Alice
  can calculate other probability distributions from it,
  synchronically, including distributions for other informationally
  complete measurements which she might carry out in the distant
  future.}

Individual statements and arguments drawn from the writings of other
scientists can sometimes fit neatly within the QBist programme.
Examples come to mind in the works of Aaronson~\cite[pp.\ xii--xiii,
  110]{Aaronson2013}, Bacon~\cite{Bacon2005}, Baez~\cite{Baez2003},
Bell~\cite{Mermin-Vienna, Mermin-Bell}, Einstein~\cite{Fuchs2014b},
Feynman~\cite[p.\ 6\,-7]{FeynmanLP}, Nielsen~\cite{Nielsen2004b},
Peierls~\cite{Mermin-Bell}, Schr\"odinger~\cite{Mermin14,
  Mermin-Vienna, Mermin-Bell} and others.  This is not to claim that
any of these authors are QBist or proto-QBist (the latter term being
also unpleasantly teleological).  Indeed, one can find self-identified
non-QBists and critics of QBism who agree with QBists on nontrivial
points~\cite{Moldoveanu2015, Brukner2015}.  Physicists and their
opinions are sufficiently complicated that we cannot pigeonhole them
based on isolated snippets of text.  Placement and classification
require more systematic study than that, if they are to have any
meaning.  With this concern in mind, we turn to surveying the writings
of von Neumann.  First, we shall see that von Neumann's interpretation
of probability, though it displayed varying nuances over time, never
aligned with that advocated by Fuchs, Mermin and Schack.  Then, we
will study the evidence indicating that von Neumann made a category
distinction between different kinds of quantum states that QBists (and
some other varieties of Quantum Bayesians) do not.  Having established
that ``Quantum Bayesian'' is not a good description for von Neumann's
thought, we will turn to the argument that underlies the claim in
Wikipedia.  Digging into the material that ostensibly supports that
claim will reveal that support to be rather insubstantial.

\section{Von Neumann on Probability}

\subsection{Frequentism (1932)}

To begin with, we examine von Neumann's \booktitle{Mathematical
  Foundations of Quantum Mechanics,} hereinafter \MFQM.  This book is
indicative of von Neumann's thinking in 1932, the time of the
publication of the original German edition.  Quoting from \MFQM, page
298:
\begin{quotation}
However, the investigation of the physical quantities related to a
single object $S$ is not the only thing which can be done --
especially if doubts exist relative to the simultaneous measurability
of several quantities.  In such cases it is also possible to observe
great statistical ensembles which consist of many systems
$S_1,\ldots,S_N$ (i.e., $N$ models of $S$, $N$
large).\textsuperscript{156}
\end{quotation}
Note 156 reads as follows:
\begin{quotation}
Such ensembles, called collectives, are in general necessary for
establishing probability theory as the theory of frequencies.  They
were introduced by R.\ v.\ Mises, who discovered their meaning for
probability theory, and who built up a complete theory on this
foundation (cf., for example, his book, ``Wahrscheinlichkeit,
Statistik and ihre Wahrenheit,'' Berlin, 1928).

[Solche Gesamtheiten, Kollektive gennant, sind \"uberhaupt notwendig
  um die Wahrscheinlichkeitsrechnung als Lehre von den H\"aufigkeiten
  begr\"unden zu k\"onnen.  Sie wurden von R.\ v.\ \textsc{Mises}
  eingef\"uhrt, der ihre Bedeutung f\"ur die
  Wahrscheinlichkeitsrechnung erkannte, und einen entsprechenden
  Aufbau derselben durchf\"uhrte (vgl.\ z.\ B.\ sein Buch
  Wahrscheinlichkeit, Statistik und ihre Wahrheit, Berlin 1928).]
\end{quotation}
Notice the discrepancy in the titles given for von Mises' book;
apparently, the translator made an error here.  In fact, von Neumann
also errs in this passage, as the ``ihre'' is an interpolation.
Nevertheless, the meaning of the passage is clear: in 1932, von
Neumann interpreted probability in a frequentist manner.

Evidence of this occurs throughout \MFQM, in fact.  For example,
``expectation value'' is defined as ``the arithmetic mean of all
results of measurement in a sufficiently large statistical ensemble''
(p.\ 308).  L\"uders, who improved upon von Neumann's theory of
measurement~\cite{Lueders1951}, also thought in terms of ``an ensemble
of identical and independent systems'' [``einer Gesamtheit
  gleichartiger und unabh\"angiger Systeme''].  This is one example of
later researchers not finding a Bayesian message in von Neumann.

\subsection{Probability as Extended Logic (c. 1937)}

To see how von Neumann's thinking on the foundations of probability
changed, we turn next to an unfinished manuscript from about 1937,
which is included in his \booktitle{Collected Works}~\cite{VN-CW}.  Von
Neumann imagines a collection of a large number of ``specimens'' of a
physical system $S_1$ and considers interpreting the transition
probability $P(\mathfrak{a},\mathfrak{b}) = \theta$ in terms of a
relative frequency:
\begin{quotation}
\noindent [I]f we measure on each $S_1^*,\ldots,S_N^*$ first
$\mathfrak{a}$, and then in immediate succession $\mathfrak{b}$, and
if then the number of those among $S_1^*,\ldots,S_N^*$ where
$\mathfrak{a}$ is found to be true is $M$, and the number of those
where $\mathfrak{a}$, $\mathfrak{b}$ are both found to be true is
$M'$, then:

(H) $P(\mathfrak{a}, \mathfrak{b}) = \theta$ means that $M'/M \to
\theta$ for $N \to \infty$.

This view, the so called ``\emph{frequency theory of probability}''
has been very brilliantly upheld and expounded by R.\ V.\ Mises.  This
view, however, is not acceptable to us, at least not in the present
``logical'' context.  We do not think that (H) really expresses a
convergence-statement in the strict mathematical sense of the
word---at least not without extending the physical terminology and
ideology to infinite systems (namely, to the entirety of an infinite
sequence $S_1^*, S_2^*, \ldots$)---and we are not prepared to carry
out such an extension at this stage.  The approximative forms of~(H),
on the other hand, are mere probability-statements, e.g. ``Bernoulli's
law of great numbers'' [\ldots\!] And such probability-statements are
again of the same nature as the relation $P(\mathfrak{a},\mathfrak{b})
= \theta$, which they should interpret.
\end{quotation} 
Von Neumann then makes the following declaration:
\begin{quotation}
We prefer, therefore, to disclaim any intention to interpret the
relations $P(\mathfrak{a},\mathfrak{b}) = \theta$ ($0 < \theta < 1$)
in terms of strict logics.  In other words, we admit:

\emph{Probability logics cannot be reduced to strict logics, but
  constitute an essentially wider system than the latter, and
  statements of the form $P(a,b) = \theta$ ($0 < \theta < 1$) are
  perfectly new and \emph{sui generis} aspects of physical reality.}

So probability logics appear as an essential extension of strict
logics.  This view, the so-called ``logical theory of probability'' is
the foundation of J.\ N.\ [sic] Keynes's work on the subject.
\end{quotation}

In short, the later von Neumann interprets quantum probabilities as
\emph{logical probabilities.}  Moreover, he explicitly identifies this
view with that worked out by Keynes.

At this point, it is a good idea to compare Keynes' ``logical
probability'' to the thinking of F.\ P.\ Ramsey, whose interpretation
is closer to that invoked in QBism~\cite[pp.\ ix, 1225--29,
  1374]{Fuchs2014}.  Fortunately, we have a statement by Keynes
himself on this subject.  In October 1931---after Ramsey's death at
the age of twenty-six---Keynes wrote the following~\cite{Keynes1931}.
\begin{quotation}
Formal logic is concerned with nothing but the rules of
\emph{consistent} thought.  But in addition to this we have certain
``useful mental habits'' for handling the material with which we are
supplied by our perceptions and by our memory and perhaps in other
ways, and so arriving at or towards truth; and the analysis of such
habits is also a sort of logic.  The application of these ideas to the
logic of probability is very fruitful.  Ramsey argues, as against the
view which I had put forward, that probability is concerned not with
objective relations between propositions but (in some sense) with
degrees of belief, and he succeeds in showing that the calculus of
probabilities simply amounts to a set of rules for ensuring that the
system of degrees of belief which we hold shall be a \emph{consistent}
system.  Thus the calculus of probabilities belongs to formal logic.
But the basis of our degrees of belief---or the \emph{a priori,} as
they used to be called---is part of our human outfit, perhaps given us
merely by natural selection, analogous to our perceptions and our
memories rather than to formal logic.
\end{quotation}
And, having made this comparison, Keynes goes on to say,
\begin{quotation}
\noindent So far I yield to Ramsey---I think he is right.  But in
attempting to distinguish ``rational'' degrees of belief from belief
in general he was not yet, I think, quite successful.  It is not
getting to the bottom of the principle of induction merely to say that
it is a useful mental habit.  Yet in attempting to distinguish a
``human'' logic from formal logic on the one hand and descriptive
psychology on the other, Ramsey may have been pointing the way to the
next field of study when formal logic has been put into good order and
its highly limited scope properly defined.
\end{quotation}

\subsection{Debating Bohr in Warsaw (1938)}

In 1938, von Neumann attended a conference in Warsaw on ``New Theories
in Physics.''  The meeting, which ran from 30 May to 3 June, was
attended by Bohr, Brillouin, de Broglie, C.\ G.\ Darwin, Eddington,
Gamow, Kramers, Langevin, Wigner and others.  Bohr presented a report
on ``The Causality Problem in Atomic Physics,'' to which von Neumann
replied in the discussion afterward~\cite{VN-WARSAW}.  Von Neumann's
remarks begin by interpreting probabilities in terms of ensembles:
\begin{quotation}
If we wish to analyse the meaning of the statistical statements of
quantum mechanics, we must necessarily deal with \ogm
ensembles\cgm\ of a great number of identical systems, and not with
individual systems.
\end{quotation}
He segues, however, into a discussion of quantum logic, arguing that
the central point is the failure of the distributive law.  This leads
to the following:
\begin{quotation}
A complete derivation of quantum mechanics is only possible if the
propositional calculus of logics is so extended, as to include
probabilities, in harmony with the ideas of J.\ M.\ Keynes.  In the
quantum mechanical terminology : the notion of a \ogm transition
probability\cgm\ from~$a$ to~$b$, to be denoted by~$P(a,b)$ must be
introduced.  ($P(a,b)$ is the probability of~$b$, if $a$ is known to
be true.  $P(a,b)$ can be used to define $a \leqq b$ and $-a$ : $P(a,b)
= 1$ means $a \leqq b$, $P(a,b) = 0$ means $a \leqq -b$.  But $P(a,b) =
\phi$, with a $\phi > 0$, $< 1$ is a new \ogm sui
generis\cgm\ statement, only understandable in terms of
probabilities.)
\end{quotation}

It is interesting that von Neumann does not attempt to use Keynesian
logical-probability theory to define \emph{single-shot}
probabilities.  Instead, he still treats statistical statements as
having meaning only for ensembles.

\subsection{Game Theory (1944)}

Von Neumann coauthored the textbook \booktitle{Theory of Games and
  Economic Behavior} with Oskar Morgenstern.  The book, first
published in 1944, is frequentist in orientation, though the authors
express this as a matter of convenience rather than necessity.  Von
Neumann and Morgenstern call the ``interpretation of probability as
frequency in long runs'' a ``perfectly well founded'' notion, but they
leave the door open to alternative conceptions of
probability~\cite{VN-GAME}.  Morgenstern later
explained~\cite{Morgenstern1976},
\begin{quotation}
\noindent We were, of course, aware of the difficulty with the logical
foundations of probability theory.  We decided we would base our
arguments on the classical frequency definition of probability, but we
included a footnote saying that one could axiomatize utility and
probability together and introduce a subjective notion of
probability.  This was done later by others.
\end{quotation}
For the work in question, see Pfanzagl~\cite{Pfanzagl1959,
  Pfanzagl1967, Pfanzagl1971}.

\subsection{Generating Random Numbers (1951)}

One of von Neumann's memorable remarks has gained a certain infamy:
``Any one who considers arithmetical methods of producing random
digits is, of course, in a state of sin.''  This quotation occurs in
an item in a 1951 volume of conference proceedings, where von Neumann
also discusses physical phenomena which can be used to generate random
numerical sequences~\cite{VN-RAND}.  He proposes ``nuclear accidents''
as the ideal source, which in the era after Chernobyl and Fukushima
comes across as slightly ominous.  However, in context it is plain
enough that the ``accidents'' in question are events like individual
clicks from a Geiger counter.
\begin{quotation}
\noindent There are nuclear accidents, for example, which are the
ideal of randomness, and up to a certain accuracy you can count them.
One difficulty is that one is never quite sure what is the probability
of occurrence of the nuclear accident.  This difficulty has been
overcome by taking larger counts than one [does] in testing for either
even or odd.  To cite a human example, for simplicity, in tossing a
coin it is probably easier to make two consecutive tosses independent
than to toss heads with probability exactly one-half.  If independence
of probability tosses is assumed, we can reconstruct a 50--50 chance
out of even a badly biased coin by tossing twice.  If we get
heads-heads or tails-tails, we reject the tosses and try again.  If we
get heads-tails (or tails-heads), we accept the result as heads (or
tails).  The resulting process is rigorously unbiased, although the
amended process is at most 25 percent as efficient as ordinary
coin-tossing.
\end{quotation}

The language here is \emph{prima facie} frequentist or
propensity-inclined, treating probabilities as unknown quantities to
be measured (``one is never quite sure what is the probability of
occurrence of the nuclear accident'').  A Bayesian can give meaning to
statements about ``unknown probabilities''---this is the territory of
the de Finetti theorem~\cite{CFS2001}---but von Neumann's phrasing does not
sound like a stringent Bayesian's first choice of words.

In summary, von Neumann's interpretation of probability moved from a
kollectiv-based strict frequentism to a Keynesian view.  Nowhere do we
find an outright endorsement of personalist Bayesianism; the closest
approach is much later than \MFQM, is not in the context of quantum
theory, and is itself mixed in with a claim that thinking of
probability as long-run frequency is good enough for practical
purposes.

\section{Pure and Mixed States}

Von Neumann's philosophy of probability, and the way his thinking
changed over time, has been discussed by others---for example, by
Bub~\cite{Bub1979}, R\'edei~\cite{Redei2001}, Stairs~\cite{Stairs1985}
and Valente~\cite{Valente2009}.  Less remarked-upon, but also
important to this comment, are von Neumann's statements concerning the
distinction between \emph{pure} and \emph{mixed} quantum states.

Returning to \MFQM, on page 295 we find the following:
\begin{quotation}
In the state $\phi$ the quantity $\mathfrak{R}$ has the expectation
value $\rho = (R\phi, \phi)$ and has as its dispersion $\epsilon^2$
the expectation value of the quantity $(\mathfrak{R} - \rho)^2$, i.e.,
$((R- \rho\cdot 1)^2\phi, \phi) = ||R\phi||^2 - (R\phi, \phi)^2$
(cf.\ Note 130; all these are calculated with the aid of $\bar{\rm
  E}\cdot$!)\ which is in general $> 0 $ (and $= 0$ only for $R\phi =
\rho\cdot\phi$, cf.\ III.3.) -- therefore there exists a statistical
distribution of $\mathfrak{R}$, even though $\phi$ is one individual
state -- as we have repeatedly noted.)  But the statistical character
may become even more prominent, if we do not even know what state is
actually present -- for example, when several states
$\phi_1,\phi_2,\ldots$ with the respective probabilities
$w_1,w_2,\ldots$ ($w_1\geq 0$, $w_2\geq 0$,$\ldots$, $w_1+w_2+\cdots =
1$) constitute the description.  Then the expectation value of the
quantity $\mathfrak{R}$, in the sense of the generally valid rules of
the calculus of probabilities is $\rho' = \sum_n
w_n\cdot(R\phi_n,\phi_n)$.
\end{quotation}
The language here indicates that for von Neumann, a pure state is
something an individual system has, and a more general density matrix
stands for an ensemble in which different pure states are physically
present with different frequencies.

Von Neumann writes freely of properties possessed by quantum systems
(p.\ 338):
\begin{quotation}
\noindent Instead of saying that several results of measurement (on
$S$) are known, we can also say that $S$ was examined in relation to a
certain property $\mathfrak{E}$ and its presence was
ascertained. [\ldots\!] The information about $S$ therefore always
amounts to the presence of a certain property $\mathfrak{E}$ which is
formally characterized by stating the projection $E$.
\end{quotation}
And, shortly thereafter, bluntly:
\begin{quotation}
\noindent That is, if $\mathfrak{E}$ is present, the state is $\phi$.
\end{quotation}
An exhaustive measurement fixes the value of a physical property, and
the presence of a physical property can mandate the correctness of a
choice of quantum state.  Unsharp measurements, in von Neumann's
development, ``are incomplete and do not succeed in determining a
unique state'' (p.\ 340).  Again, we see the language making a
category distinction between quantities which are ``states'' and more
general entities which are not.  Later, von Neumann writes that a
density operator which is a ``mixture of several states'' is ``not a
state'' itself (p.\ 350).  This distinction is maintained throughout
his discussion of what we now call the von Neumann entropy.

This idea, that pure and mixed states are qualitatively different
kinds of entity, went unquestioned by Bohm~\cite{Bohm1951}, but was soon
challenged by Jaynes~\cite{Jaynes1957}.  One can conceive of
uncertainty intrinsic to a pure state and uncertainty about which pure
state might be present.  However, as Jaynes writes,
\begin{quotation}
\noindent If the former probabilities are interpreted in the objective
sense, while the latter are clearly subjective, we have a very
puzzling situation.  Many different arrays, representing different
combinations of subjective and objective aspects, all lead to the same
density matrix, and thus to the same predictions.
\end{quotation}
This argument was later made by Ochs \cite{Ochs1981}, and by Caves,
Fuchs and Schack~\cite{CFS2001}.  The latter authors write, ``a mixed
state has infinitely many ensemble decompositions into pure
states''---even into \emph{different numbers of}\!\ pure states---``so
the distinction between subjective and objective becomes hopelessly
blurred.''\footnote{The mathematical point of the multiplicity of
  ensemble decompositions was made most famously by Hughston, Jozsa
  and Wootters~\cite{Hughston1993} in 1993.  It was also demonstrated
  almost sixty years earlier by Schr\"odinger~\cite{Schroedinger1936},
  who disclaimed priority for the result, suggesting that some form of
  the idea was folk knowledge or shared conversationally at the time.
  The fact that a mixed state can be decomposed in multiple ways is
  one of the phenomena reproduced in the ``epistricted'' models of
  Spekkens \emph{et al.}\ \cite{Spekkens2007, Spekkens2014}.}

Arguably, it would be more consistent for a strict Kollectivist to
treat all quantum states, pure and mixed, in ensemble terms, where
pure states correspond to the most purified possible ensembles.
However, von Neumann's statements about the presence of physical
properties determining unique states clash with this position.

\section{Measurement and Subjectivity}

As we noted in Section~\ref{sec:qbism}, it is not so difficult to find
in physicists' writings statements which, taken in isolation, are
compatible with QBism.  The more important question is whether a
verbal corpus yields up enough of these cherries to fill a bowl.

The last chapter of \MFQM\ concerns ``The Measuring Process.''  Here,
we find mentions of ``subjective perception'' and ``the intellectual
inner life of the individual'' (p.\ 418).  Surely this is where we
should look for evidence of von Neumann's Quantum-Bayesian
sympathies.  

He writes (p.\ 420),
\begin{quotation}
\noindent Indeed experience only makes statements of this type: an
observer has made a certain (subjective) observation; and never any
like this: a physical quantity has a certain value.
\end{quotation}
But the idea that we ultimately rely on sense impressions to
adjudicate between scientific models is hardly original to quantum
mechanics, or to Bayesian interpretations thereof.  A form of the idea
is attributed to Democritus~\cite{DK68B9}, and Lucretius discussed it
in verse~\cite{Lucretius}.  Schr\"odinger commented, ``Quantum
mechanics forbids statements about what really exists---statements
about the object. Its statements deal only with the object-subject
relation.''  However, he continued, ``this holds, after all, for any
description of nature''; the crucial point is that in quantum physics,
it ``holds in a much more radical and far reaching sense''~\cite{AJP}.

In classical mechanics, the mass of an object is a basic property
which that object has whether or not a physicist is nearby to be
interested in it.  Let us imagine a physicist working in the days
before quantum theory.  He drops a rock on his foot and sees red---a
subjective perception.  He then uses this perception, which is part of
his ``inner life,'' to rule out the hypothesis that the rock is of
negligible mass.  The variable $m$ in his equations refers to an
intrinsic property of the rock, not to any of his sensations, even
though his sensory perceptions are what he uses to assign a value (or
a spread of reasonable values) to the variable $m$.  For the
Newtonian, the rock \emph{has a mass,} although the Newtonian might
admit when pressed that any \emph{particular value used in a
  calculation} is chosen because it adequately summarizes past
experiences and helps to predict future ones.

The critical question is whether von Neumann reads the mathematical
entities appearing in the quantum formalism as physical quantities
akin to Newtonian masses, about which we can use subjective
perceptions to make estimations, or if he reads those mathematical
entities as standing for perceptions themselves.  The blanket
statement quoted above that ``experience only makes statements of this
type'' certainly suggests that he views the point about ``(subjective)
observation'' to be applicable to classical physics.  The evidence we
saw in the previous section indicates that von Neumann treats quantum
states as physical properties held by objects themselves, that is, as
more analogous to the mass of a Newtonian rock than to experiences in
the flow of ``intellectual inner life.''

At one point in the Warsaw proceedings, von Neumann does approach a
statement which might not sound completely out of place coming from a
QBist.  In a later discussion than the exchange we examined above, the
Warsaw proceedings record the following~\cite[p.\ 44]{VN-WARSAW}.
\begin{quotation}
\textsc{Professor von Neumann} thought that there must always be an
observer somewhere in a system : it was therefore necessary to
establish a limit between the observed and the observer.  But it was
by no means necessary that this limit should coincide with the
geometrical limits of the physical body of the individual who
observes.  We could quite well \ogm contract\cgm\ the observer or \ogm
expand\cgm\ him : we could include all that passed within the eye of
the observer in the \ogm observed\cgm\ part of the system --- which is
described in a quantum manner.  Then the \ogm observer\cgm\ would
begin behind the retina.  Or we could include part of the apparatus
which we used in the physical observation --- a microscope for
instance --- in the \ogm observer\cgm.  The principle of \ogm
psycho-physical parallelism\cgm\ expresses this exactly : that this
limit may be displaced, in principle at least, as much as we wish
inside the physical body of the individual who observes.  There is
thus no part of the system which is essentially the observer, but in
order to formulate quantum theory, an observer must always be placed
somewhere.
\end{quotation}

Terms like ``observer'' and ``measurement'' imply an essential
passivity, Fuchs and Schack have argued~\cite{RMP}; such words suggest
a casual, uninvolved reading-off, rather than a participatory act.
But if we replace ``observer'' with ``agent'' in von Neumann's
concluding line, we would have the statement, ``In order to formulate
quantum theory, an agent must always be placed somewhere'';
\emph{this} claim, that \emph{agent} is a fundamental concept which
quantum theory is built upon, would fit within QBism.

To a QBist, this is the killing flaw in von Neumann's interpretation
of quantum mechanics.  On the one hand, von Neumann affirms that one
cannot formulate the theory without an observer, but on the other,
quantum states are physical properties of systems outside the
observer, and probabilities are frequencies in kollectivs or Keynesian
logical valuations---conceptions of probability which try to delete
the agent at all cost.\footnote{Arguably, von Neumann is not always
  consistent in his treatment of ``psycho-physical parallelism.''  A
  careful reading of \MFQM{} \S VI.3 suggests that it elides the
  ``limit between the observed and the observer'' which he deems
  essential, both in \MFQM{} (\S VI.1, p.\ 420) and at the Warsaw
  conference.  But teasing out the meaning of ``psycho-physical
  parallelism'' is no simple task~\cite{HayPeres1998}, and for the
  purposes of this essay, pursuing it in greater depth is not
  essential.}

In broad overview, von Neumann's approach to quantum measurement
begins with a physical system, which then interacts with some kind of
measuring apparatus, which is then studied by an observer.  The
system-apparatus and apparatus-observer interactions are treated as
physically distinct kinds of time evolution.  Von Neumann calls the
difference between these processes ``very fundamental'' (\S VI.1,
p.\ 418).

If one takes quantum states to be intellectual tools held by an
individual agent, then the procedure of inserting an intermediate
apparatus adds nothing to the basic philosophical understanding of
what quantum theory is about.  It could well be a beneficial
mathematical exercise, part of working out what to do when making use
of multipartite systems.  (There are good practical reasons to
understand how the quantum formalism applies to a system one of whose
parts is a probe or an ancilla for the other, or is a communication
channel whose function is limited in some way.)  Regardless, if quantum
states are personalist Bayesian quantities, then introducing probes
and ancillas brings nothing intrinsically new.  And if \emph{von
  Neumann had seen quantum states in anything like the QBist fashion,}
it is difficult to find a rationale for why \MFQM's entire chapter on
``the measuring process'' assumes the shape it does.

Fuchs has written, ``von Neumann's setting the issue of measurement in
these terms was the great original sin of the quantum foundational
debate''~\cite[p.\ 2035]{Fuchs2014}.

\section{Measurement Redux: The ``Quantum Bayes Rule''}

Wikipedia attributes the statement ``The first quantum Bayesian was
von Neumann'' to R.\ F.\ Streater~\cite{Streater2007}.  As mentioned
earlier, the effect of saying this in an article which primarily
concerns QBism is to claim that von Neumann was himself either QBist
or something much like it.  Looking up this source, we find that what
Streater calls ``Quantum Bayesianism'' could indeed reasonably include
QBism.  For example, he states that Bayesians ``attribute \emph{all}
the entropy in a state to the lack of information in the observer''
(p.\ 71).  And in discussing density matrices, he writes, ``the
Bayesian's $\rho$ is entirely about his knowledge'' (p.\ 72).  So, the
meaning of the statement created by placing it within the Wikipedia
article is not too much of a stretch.

Streater bases his claim that von Neumann was the ``first quantum
Bayesian'' on \MFQM, never addressing the plainly frequentist
orientation of that book.  Nor does Streater refer to the passages
about ``subjective perception'' and ``experience.''

The root of the confusion appears to be that, towards the end of
\MFQM, von Neumann derives a formula which turns out in retrospect to
be a specialized case of a quantum analogue of the Bayes conditioning
rule.  Von Neumann motivates his argument with the following
(p.\ 337):
\begin{quotation}
If anterior measurements do not suffice to determine the present state
uniquely, then we may still be able to infer from those measurements,
under certain circumstances, with what probabilities particular states
are present.  (This  holds in causal theories, for example, in
classical mechanics, as well as in quantum mechanics.)  The proper
problem is then this:  Given certain results of measurements, find a
mixture whose statistics are the same as those which we shall expect
for a system {\bf S} of which we know only that these measurements
were carried out on it and that they had the results mentioned.
\end{quotation}
Here, von Neumann treats quantum states as analogous to the
\emph{physical} states of classical mechanics, \emph{i.e.,} to points
in phase space.  In classical mechanics, if a system could be at one
of multiple points in its phase space, we write a Liouville
probability density over that space; we can infer that for von
Neumann, it is \emph{mixtures} which are analogous to Liouville
densities.  This is in sharp contrast with QBist and much other
Quantum-Bayesian thinking, in which \emph{all} quantum states, however
pure, are expressions of an agent's probability assignments.

As always in \MFQM, a system has a state, even if we don't know what
that state is.  And, as always in \MFQM, statistics means ensembles of
identically prepared systems:
\begin{quotation}
\noindent If, for many systems ${\bf S}'_1,\ldots,{\bf S}'_M$
(replicas of~{\bf S}), these measurements give the results mentioned,
then this ensemble $[{\bf S}'_1,\ldots,{\bf S}'_M]$ coincides in all
its statistical properties with the mixture that corresponds to the
results of the measurements.
\end{quotation}
Changes in statistical properties mean the creation of new ensembles
with different population demographics:
\begin{quotation}
That the results of the measurements are the same for all ${\bf
  S}'_1,\ldots,{\bf S}'_M$ can be attributed, by {\bf M.}, to the fact
that originally a large ensemble $[{\bf S}_1,\ldots,{\bf S}_N]$ was
given in which the measurements were carried out, and then those
elements for which the desired results occurred were collected into a
new ensemble.  This is then $[{\bf S}'_1,\ldots,{\bf S}'_M]$.
\end{quotation}
Here, {\bf M.}\ refers to the measurement postulate, ``If the physical
quantity $\mathfrak{R}$ is measured twice in succession in a system
{\bf S}, then we get the same value each time'' (p.\ 335).

So, we have \emph{something like} the updating of probabilities by the
Bayes rule.  However, von Neumann phrases the scenario in completely
Kollectivist language.  Merely invoking Bayes' theorem does not make
one a Bayesian.  For example, von Mises makes use of Bayes' theorem,
calling it ``a proposition applying to an infinite number of
experiments,'' or in other words, to a
kollectiv~\cite[p.\ 123]{Mises1957}.  One could be wholly agnostic
about the interpretation of probability, setting up the theory from
measure-theoretic or abstract-algebraic axioms~\cite{Tao2012};
multiplication and division of probabilities would then be legitimate
operations having meaning only with respect to those axioms.

It is in this context that von Neumann mentions ``a priori'' and ``a
posteriori'' probabilities.  These terms could be glossed in a
Bayesian way, but only at the cost of ignoring everything else \MFQM{}
says about the interpretation of probability, including the discussion
of ``ensembles'' in the same paragraph.  And even if one were to do
so, the way in which von Neumann allows pre-existing physical
properties to determine quantum states would imply a view in which
mixed states are \emph{Bayesian probability distributions over pure
  states.}  (And though it could potentially be called ``quantum
Bayesian,'' it is definitely not QBist.)  This is a difficult position
to maintain, per the Jaynesian argument given above.  Furthermore,
given the professed Kollectivism of \MFQM, we should recall what von
Mises said about these terms~\cite[p.\ 46]{Mises1957}:
\begin{quotation}
It is useful to introduce distinct names for the two probabilities of
the same attribute, the given probability in the initial collective
and the calculated one in the new collective formed by partition.  The
current expressions for these two probabilities are not very
satisfactory, although I cannot deny that they are impressive enough.
The usual way is to call the probability in the initial collective the
\emph{a priori,} and that in the derived collective the \emph{a
  posteriori} probability.
\end{quotation}
This usage exactly parallels von Neumann's.  To continue:
\begin{quotation}
\noindent The fact that these expressions suggest a connexion with a
well-known philosophical terminology is their first deficiency in my
eyes.  Another one is that these same expressions, \emph{a priori} and
\emph{a posteriori,} are used in the classical theory of probability
in a different sense as well, namely, to distinguish between
probabilities derived from empirical data and those assumed on the
basis of some hypothesis; such a distinction is not pertinent in our
theory.  I prefer, therefore, to give to the two probabilities less
pretentious names, which have less far-reaching and general
associations.  I will speak of \emph{initial probability} and
\emph{final probability,} meaning by the first term the probability in
the original collective, and by the second one, the probability (of
the same attribute) in the collective derived by partition.
\end{quotation}
Von Neumann uses the more common terminology, but the meaning which
\MFQM{} vests in the words is, by all evidence, the same as that which
von Mises does.  The result is not an argument that probabilities
should be seen as quantified fervencies of belief, but rather that a
certain problem involving ensemble frequencies admits nonunique
solutions.

Von Neumann derives his basic relation betwen initial and final
ensembles by considering the following procedure (p.\ 340).  We
measure some binary physical property $\mathfrak{E}$ on each element of the
initial ensemble, which is described by the statistical operator $U_0$.
The elements for which this measurement yields the outcome 1 (instead
of~0) are collected to form a new ensemble, whose statistical operator
is $U$.  The two ensembles are related by
\begin{equation}
U = \sum_n (U_0 \phi_n, \phi_n) P_{[\phi_n]}.
\end{equation}
Here, $P_{[\phi_n]}$ is the projector onto the state $\phi_n$, and the
set $\{\phi_1,\phi_2,\ldots,\phi_n\}$ is an orthonormal basis which
spans the subspace in which measuring $\mathfrak{E}$ yields the value 1.

L\"uders criticized von Neumann's result and proposed a
correction~\cite{Lueders1951}.  A more modern way to represent
state-change upon measurement is to write the L\"uders rule using the
mathematics of \emph{effects and operations.}  A measurement is a
positive operator valued measure (POVM) which furnishes a resolution
of the identity:
\begin{equation}
\sum_k E_k = 1,
\end{equation}
and each of the $\{E_k\}$ can be written
\begin{equation}
E_k = \sum_l A^\dag_{kl} A_{kl}.
\end{equation}
Here, the index $k$ labels the possible outcomes of the measurement.
If the initial density operator is $U_0$, then upon obtaining the
outcome $k$, we update the density operator to
\begin{equation}
U = \frac{\sum_l A_{kl} U_0 A^\dag_{kl}}
         {\tr(E_k U_0)}.
\end{equation}
For analyses of how this update rule is analogous to, or a variant of,
Bayesian conditioning, see Schack, Brun and Caves~\cite{Schack2001};
and also Fuchs~\cite{Fuchs2002}.  Illustrative examples are developed
in Fuchs and Schack~\cite{Fuchs2009}.

Streater bases his criticism of von Neumann on that found in the
textbook of Krylov~\cite{Krylov1979}, writing, ``Krylov did not
believe that all the characteristics of the state reflect only the
lack of knowledge of the observer, but that there was a physical
state, $\rho_0$ out there to be found.''  But this is exactly what von
Neumann stated: recall that for him, a ``mixture of several states''
is ``not a state'' itself (\MFQM, p.\ 350).

Krylov's primary complaint with von Neumann (pp.\ 184--85) is the
multiplicity of valid decompositions of mixed-state density operators.
\begin{quotation}
Firstly, in using the statistical operator, we assume the selection of
a certain orthogonal system of coordinates in the subspace delimited
by an inexhaustively complete experiment and we also assume a certain
choice of weights $w_i$.  [\ldots\!]  A change in the orthogonal
system means, generally speaking, a transition to another physical
state described by the statistical operator, to what is said to be
another statistical aggregate.  (``A state'' is understood here in a
more general sense than a state exhaustively completely determined,
using a $\Psi$-function.)  [\ldots\!]  Therefore, the selection of a
certain orthogonal system of functions and the fixing of certain
weights $w_i$, which the von Neumann operator presupposes, amounts to
the introduction of some {\sl physically fictitious} properties of the
reality being described.
\end{quotation}
That is, Krylov finds the conjunction of the following two statements
unacceptable:
\begin{itemize}
\item A pure quantum state is a physical property of a system.
\item The quantum formalism implies that a mixed-state statistical
  operator has multiple decompositions into linear combinations of
  pure states.
\end{itemize}
Krylov insists upon the former and therefore finds the latter
unsatisfactory.  As we discussed earlier, QBists (and some other
varieties of quantum Bayesians) agree that these two statements clash
with each other, but discard the first instead.  Von Neumann holds
onto both (\MFQM, \S IV.3).

\section{Discussion}

We have seen that ``Quantum Bayesian'' is not at all a good
description of von Neumann.  Along the way, we have encountered some
of the practical issues that make classifying scientists a difficult
problem.  It is worth considering these issues more generally.  Having
done so, we will conclude with some contemplations about the platform
that made the claim ``the first Quantum Bayesian was von Neumann''
visible enough to be noticed in the first place.

A good classification summarizes, as succinctly as possible, the known
statements and actions of an individual, and should have predictive
power for statements yet unmade or undiscovered.  Furthermore, the
descriptive terms in commonest circulation should provide the most
useful and meaningful understanding of relevant distinctions that
painting with a broad brush can convey~\cite{Brown2010}.  It is, to
say the least, debatable that the jargon we have today on the
philosophical side of quantum theory fares at all well in this regard.
Peres quips, ``There seems to be at least as many Copenhagen
interpretations as people who use that term, perhaps even
more''~\cite{Peres2002}.  For \.{Z}ukowski, Copenhagen is ``different
for every apostle''~\cite{Zukowski2015}.  Indeed, a
good case can be made that the idea of a unified ``Copenhagen
interpretation'' was a myth of the 1950s, which now elides in our
perception the differences among views held by physicists thirty years
earlier~\cite{Howard2004, Camilleri2009, Camilleri2015}.  Similarly,
Kent tabulates at least twenty-one varieties of Everettian
interpretation~\cite{Kent2010}.  These can differ in underacknowledged
ways from Everett's original~\cite{Kent2011}, and many are ``generally
incompatible'' with one another~\cite{Kent2014}.

The terms we employ are signifiers that we use to determine what we
bother to read.  They establish the distinctions which we allow to
exist between the players in the histories we retell.  Then we pose
those players according to our sentiments for the person and for the
philosophy.\footnote{I myself have heard Feynman claimed as a
  Copenhagener, a Gell-Mannian, an Everettic and an enthusiast for
  nonlocal hidden variables.  The implication is that the correct
  interpretation of the quantum will be decided, not even by opinion
  poll~\cite{Nielsen2004, Fuchs2014b}, but by who occupies the most
  shelf space in the Caltech bookstore.}

The story of von Neumann exemplifies other challenges as well.  We
scientists change our views over time.  We leave fragmentary records
of our thoughts, not atypically muddled by the compromises of
coauthorship and journal publication.  (The rise of an electronic
preprint culture has, among other things, provided a way to track what
squeezing our works into journals can do to them; for an example,
see~\cite{Ferrie2014}.)  And because our attitudes can be moving
targets, combining a physicist's statement from year $N$ with another
statement they made in year $N + 20$ to deduce what they ``must
logically have believed'' is an exercise fraught with a scholarly kind
of peril.

The scholarly literature is quite capable of spreading urban legends
of its own~\cite{Rekdal2014}.  What happens when we bring
Wikipedia into the mix?

The Wikipedia project bills itself as ``the free encyclopedia that
anyone can edit''; even if we gloss ``anyone'' as ``anyone with an
Internet connection,'' this is only true to a first approximation.
Individual users, or the machines they edit from, can be blocked from
editing for various lengths of time for offenses like persistent hate
speech~\cite{Ferguson2014}.  Also, individual pages can be protected
from editing to different extents.  Wikipedia has its own policies and
community institutions~\cite{Paling2015}, often referred to by
acronyms (NPOV, NOR, FAC, FARC and so forth).  These, in addition to
sheer size and visibility, distinguish Wikipedia from other
applications of the wiki concept, such as the nLab~\cite{nLab}.  For
example, Wikipedia has a ``No Original Research'' policy~\cite{WPNOR},
but the nLab has as one of its primary goals the facilitation of
original mathematical work.

On Wikipedia, the ``right to figure forever in the history of the
subject like a fly in amber''~\cite{James1909} can be supported even
by a mention in nothing more substantial than the popular press, like
by \booktitle{New Scientist} magazine, despite the well-known failure
modes of that industry~\cite[p.\ 2221]{Fuchs2014}.

Thanks to the No Original Research policy mentioned earlier,
correcting misconceptions propagated by popular-science magazines and
the like cannot begin with Wikipedia itself.  The NOR policy makes
sense for what Wikipedia is and what it tries to be: an encyclopedia
is a tertiary source, rather than a primary or secondary one.
Moreover, Wikipedia lacks the infrastructure to evaluate original
scholarship, and identifying who wrote what in its articles is an
arduous task, making it a poor place to advance new claims in a
forthright way.  Academic life depends on receiving credit for one's
own work, and the way Wikipedia articles are made flattens all
contributions together, obscuring authorship.  This essay is an
attempt to do in a different venue what cannot feasibly be done within
Wikipedia alone: set the record straight.

\bigskip

I thank John DeBrota and Chris Fuchs for discussions during writing
and editing.

\appendix
\section{Things QBism Is Not}

Taking the recent writings of Fuchs, Mermin and Schack as establishing
what QBism is, we can identify some things which QBism is not.  
\begin{itemize}
\item \emph{A hidden-variable model.}  Quantum states in QBism are
  probability distributions over potential experiences, not over
  values of putative hidden variables or agent-independent ``ontic
  states.''  QBism is more compatible with the research programme
  which uses hidden-variable models to reconstruct \emph{portions} of
  quantum theory~\cite{Spekkens2007, VanEnk2007, Coecke2011,
    Coecke2011b, Bartlett2012, Spekkens2014, Ferrie2014b}.  In these
  constructions, one posits a classical theory with discrete or
  continuous degrees of freedom, and then one imposes a restriction on
  what can be known about those degrees of freedom at any one time.
  The resulting statistical theory can qualitatively reproduce
  features of quantum mechanics (no-cloning and no-broadcasting
  theorems, teleportation and so forth); in some cases, it exactly
  reproduces a subtheory of quantum mechanics, including a subset of
  the states and operations available in the full theory.  The
  phenomena \emph{not} reproduced, such as the hope of computational
  speedup, are taken as indications of what is strongly nonclassical
  about quantum physics.  The goal is to follow these hints to an
  interpretation of quantum mechanics as a statistical theory about
  something other than pedestrian hidden variables.  A QBist is not
  philosophically obligated to find this research programme appealing,
  but it is much more aligned with QBism than attempts to reproduce
  \emph{all} of quantum theory using hidden variables could be.

\item \emph{Solipsism.}  According to QBism, quantum theory concerns
  the interface between an agent who uses quantum theory and the
  external world.  Without the external world, there would be no
  interface, no subject matter for quantum theory and, indeed, no
  science~\cite{RMP, Mermin-Bell}.  Notwithstanding this, over the
  years QBism has been accused of solipsism; for examples and
  responses, see the list on pages xlv--xlvi
  of~\cite{Fuchs2014}.\footnote{A QBist could justifiably ask, ``If
    there is no world outside of my head, where do all the papers on
    Bohmian mechanics keep coming from?''}

\item \emph{``The Copenhagen Interpretation.''}  It is difficult to
  define what, historically speaking, a ``Copenhagen Interpretation''
  should be.  For the moment, it suffices to take a specific reference
  which might be designated as Copenhagenish, the quantum physics
  volume of Landau and Lifshitz~\cite{Landau1991}.  The position of
  Landau and Lifshitz differs from QBism in ways which Fuchs, Mermin
  and Schack have tabulated~\cite{AJP, Mermin-Vienna, Mermin-Bell}.
  In particular, the notions of ``classical object'' and ``classical
  `apparatus'\,'' are central to Landau and Lifshitz's interpretation,
  but not to QBism.  We should also note that Bohr and Heisenberg
  disagreed on the nature of the quantum/classical
  ``cut''~\cite{Camilleri2015}, and QBism disagrees with both of them.
\end{itemize}

The ideas just itemized are incompatible with QBism: one who holds
them is \emph{ipso facto} not a QBist.  We can also identify some
things which QBism does not mandate.  For example, a QBist does not
have to expect that human beings, when tested in psychology
laboratories, must follow a Bayesian decision-making scheme.  Fuchs,
Mermin and Schack all use a \emph{Dutch book} argument which deduces
the rules of probability from a normative requirement, and humans are
very good at falling short of normative requirements.\footnote{In
  addition, QBism does not require that the process of science as a
  whole be understood as clockwork updating in accord with Bayes'
  theorem~\cite[pp.\ 193, 500, 799, 1020, 1228, 1731]{Fuchs2014}.  The
  QBist investigations of when probabilistic coherence arguments are
  or are not in force~\cite{FuchsSchack2011, RMP} militate against
  such a view.}

On a more technical note, Fuchs and Schack demonstrate that we can
view the Born rule as an empirical addition to the bare structure of
probability theory~\cite{RMP}.  The Born rule does not derive
probabilities from some more fundamental kind of mathematical object;
instead, it relates one probability distribution to another, tying
together the probability assignments which an agent can consistly
ascribe to different experiments.

What if one began with a different bare structure, other than
probability theory as modern personalist Bayesians know it?  Following
the speculations of Greenberger and Gill~\cite[pp.\ 567,
  1265]{Fuchs2014}, we might imagine intelligent life evolving on
Jupiter; when the Jovians turn to mathematics, what constructions do
they find intuitive and compelling?  Perhaps they are filamentous
beings, and knot theory is to them as counting is to us.  When a
Jovian quantifies its beliefs and expectations, it encodes them as a
collection of knots.  Instead of a Dutch-book argument imposing
conditions on sets of real numbers, Jovian coherence is a normative
standard about the combinations of knots that a good Jovian should
strive to maintain.  Updating expectations in the light of new
experiences is a matter of undoing and retying conceptual tangles.
The Jovian art of thread manipulation is a theory of learning, helpful
in at least some corners of life, just like Bayesian probability.  And
it is physics-neutral, just as the bare structure of Terran
probability theory is.  Should the Jovians develop quantum mechanics,
the Born rule would be an empirical addition to their theory of knots,
an extra consistency or coherence condition phrased in the terms of
that theory.

Thus, the use of personalist Bayesian probability theory itself may be
secondary to a deeper physical principle; however, this line of
thought has not been developed in any detail.  And even if it were,
some conceptions of probability, such as Lewisian objective chance,
assert the existence of physical properties which are incompatible
with QBism~\cite{RMP}.  An alternative theory of learning would not
erase that incompatibility.

\section{Von Neumann, Bohm and L\"uders}

In the main text, we have shown how von Neumann's thinking, even when
it turned to notions like ``subjective inner life,'' did not align
with Quantum Bayesianism or QBism.  It is also helpful to see how
other physicists read von Neumann.  Two historically significant
examples, Bohm and L\"uders, indicate how physicists who paid close
attention to von Neumann did not take a Bayesian message from his
work.

One place where von Neumann's influence was felt is David Bohm's 1951
textbook, \booktitle{Quantum Theory}~\cite{Bohm1951}.  Bohm wrote this
book before developing what we now call ``Bohmian mechanics''; instead
of trying to make sense of quantum theory using pilot waves, it aims
to present the perceived practical orthodoxy of the time.  Bohm's
chapter on the ``quantum theory of the measurement process''
references \MFQM's discussion of that topic, with no indication that
Bohm found it flawed (p.\ 583, \S 22.1).

First, Bohm interprets quantum probabilities as long-run frequencies.
The meaning of a probability found by squaring the magnitude of a wave
function depends, he says, on ``a large number of equivalent systems''
prepared identically (p.\ 224, \S 10.30).  Furthermore, for an
individual quantum system, there is in principle a correct quantum
state; ``the physical state of the system'' determines ``the
probability of a quantum jump'' (p.\ 30, \S 2.5).  Bohm describes one
thought experiment in the following terms (p.\ 606, \S 22.11):
\begin{quotation}
\noindent The entire system, consisting of spin, $z$ co-ordinate of
the atom, apparatus which measures the $z$ co-ordinate, and apparatus
which records the results of this measurement, is assumed to have some
pure wave function when the experiment starts.  (It is not necessary
that any human observer know exactly what this wave function is.)
\end{quotation}
If one of a set of quantum states might be physically present, Bohm
treats the situation with a ``statistical ensemble of states''
(p.\ 604, \S 22.10).  Refining an ensemble to contain only a single
state ``represents absolutely no change in the state of the spin'';
indeed, Bohm goes so far as to argue that this means the terms ``mixed
state'' and ``pure state'' are misleading.
\begin{quotation}
\noindent It seems unwise to adopt a terminology that suggests the
spin changes its state (from mixed to pure) under circumstances in
which nothing changes except the observer's information about the
spin.  The phrase ``statistical ensemble of states'' provides a more
accurate description.
\end{quotation}
For Bohm and von Neumann alike, states are physically present outside
the physicist, and uncertainty about which state might be extant is
represented by statistical ensembles.  Like Bohm, von Neumann draws a
categorical line between what we would call pure states and mixed states.

L\"uders, who criticized \MFQM's treatment of state change due to
measurement, kept to \MFQM's philosophy of
probability~\cite{Lueders1951}.
\begin{quotation}
\noindent In a measurement of $R$ followed by a selection of~$r_k$,
$Z$ is transformed to
\begin{displaymath}
Z_k' = P_k Z P_k.
\end{displaymath}
$Z_k'$ is not normalized [\ldots\!]\ but instead is chosen so that the
trace shows the relative frequency of the occurrence of~$r_k$ in the
ensemble.

[Bei Messung von $R$ mit nachfolgender Aussonderung von $r_k$ geht $Z$
  \"uber in
\begin{displaymath}
Z_k' = P_k Z P_k.
\end{displaymath}
$Z_k'$ ist nicht normiert [\ldots\!]\ sondern so gew\"ahlt, da\ss{} die
Spur die relative H\"aufigkeit des Auftretens von $r_k$ in der
Gesamtheit wiedergibt.]
\end{quotation}
Here we have someone who read von Neumann carefully and offered a
correction to von Neumann's work.  L\"uders is plainly frequentist,
yet never argues that this represents a departure from von
Neumann---because, as we have seen, it doesn't.

\section{Multiplicity of Density-Matrix Decompositions}

The main text shows how the multiplicity of pure-state decompositions
of a mixed density matrix is an important point to consider.  The
mathematical derivation is a standard result~\cite{Schroedinger1936,
  Hughston1993}; however, I was unable to find a source for it online
that I did not have to pay to download.  For convenience, and to make
clear what exactly is being claimed, I reproduce the argument here.

Let $\rho$ be a density matrix, and let $\{\ket{\hat{e}_i}\}$ be a
normalized eigenbasis of~$\rho$ with eigenvalues $\{\lambda_i\}$.
Denote the rank of~$\rho$ by~$k$, and pick an arbitrary matrix $M$ of
dimensions $r \times k$, where $r \geq k$, and the columns of~$M$ are
orthonormal vectors in~$\mathbb{C}^r$.  Then, for any such choice of
matrix $M$, the set of unnormalized states
\begin{equation}
\ket{\psi_i} = \sum_{j=1}^k M_{ij} \sqrt{\lambda_j} \ket{\hat{e}_j},
\hbox{ with } i = 1,\ldots,r,
\end{equation}
provide a resolution of the density matrix $\rho$.  The proof is
straightforward.  First,
\begin{equation}
\sum_{i=1}^r \ket{\psi_i}\bra{\psi_i}
 = \sum_{i=1}^r \sum_{l,m}^k M_{il}^* M_{im} \sqrt{\lambda_l\lambda_m}
   \ket{\hat{e}_m}\ket{\hat{e}_l}.
\end{equation}
Using the orthonormality of the columns of $M$, we simplify this to
\begin{equation}
\sum_{i=1}^r \ket{\psi_i}\bra{\psi_i}
 = \sum_{m=1}^k \lambda_m \ket{\hat{e}_m}\bra{\hat{e}_m}.
\end{equation}
The quantity on the right is just the density matrix $\rho$, meaning
that
\begin{equation}
\rho = \sum_{i=1}^r \ket{\psi_i}\bra{\psi_i}.
\end{equation}
If we normalize the states $\{\ket{\psi_i}\}$ by
\begin{equation}
\ket{\hat{\psi}_i} =
 \frac{\ket{\psi_i}}{\sqrt{\braket{\psi_i}{\psi_i}}},
\end{equation}
then the density matrix is
\begin{equation}
\rho = \sum_{i=1}^r w_i \ket{\hat{\psi}_i}\bra{\hat{\psi}_i},
\hbox{ where } w_i = \braket{\psi_i}{\psi_i}.
\end{equation}
The coefficients $w_i$ give the statistical weightings of the states
in the decomposition.

\section{QBism in the Media}

This essay was originally prompted by a claim in the Wikipedia article
on ``Quantum Bayesianism,'' a page that, then as now, focused on
QBism.  Since Wikipedia relies upon secondary sources for material, it
is worth noting the other places where QBism has been treated in
popular or semi-popular venues.

QBism has been written up both in \booktitle{New
  Scientist}~\cite{Chalmers2014} and in \booktitle{Scientific
  American}~\cite{VonBaeyer2013}, though not terribly accurately in
either case, thanks to the editorial process~\cite{Mermin14b,
  Mermin-Vienna, Mermin-Bell}.  A better treatment, albeit in German,
appeared in the \booktitle{Frankfurter Allgemeine
  Sonntagszeitung}~\cite{VonRauchhaupt2014}. \booktitle{Nature}
addressed it briefly in the context of information-oriented
reconstructions of quantum theory~\cite{Ball2013}.  Later, Mermin
published in \booktitle{Nature} an opinion piece promoting
QBism~\cite{Mermin14}, which was featured on the magazine cover.

In June 2015, the pop-science website \booktitle{Quanta Magazine} ran
an interview with Fuchs~\cite{Gefter2015}.  The accompanying profile
is largely accurate, except for a figure caption that implies QBism is
a hidden-variable theory:
\begin{quotation}
\noindent A quantum particle can be in a range of possible
states. When an observer makes a measurement, she instantaneously
``collapses'' the wave function into one possible state. QBism argues
that this collapse isn't mysterious. It just reflects the updated
knowledge of the observer. She didn't know where the particle was
before the measurement. Now she does.
\end{quotation}
A better caption would go more like the following:
\begin{quotation}
\noindent In the textbook way of doing quantum physics, a quantum
particle has a ``wave function'' that changes smoothly when no one is
looking, but which makes a sharp jump or ``collapse'' when the
particle is observed. QBism argues that this collapse isn't
mysterious. It just reflects the altered expectations of the
observer. Before the measurement, she didn't know what would happen to
her when she interacted with the particle. After the measurement, she
can update her expectations for her future experiences accordingly.
\end{quotation}
(Originally, the subhead was also misleading; soon after the interview
appeared, \booktitle{Quanta} fixed the subhead, but not the figure
caption.  So it goes.)

Later, Fuchs was interviewed for the Australian Broadcasting Company's
program, \booktitle{The Philosopher's Zone}~\cite{philzone}.  A
pop-science book on QBism is also forthcoming~\cite{VonBaeyer2016}.

\section{How Wikipedia Fails, And Perhaps Why}

Wikipedia's treatment of QBism has other problems, beyond the
historical matter that the main text explores at length.  For example,
it asserts that QBism ``is very similar to the Copenhagen
interpretation that is commonly taught in textbooks.''  We can
legitimately ask what this might even mean.  As we noted above, ``the
Copenhagen interpretation'' is an ill-defined term.  Furthermore,
claiming that ``the Copenhagen interpretation'' is ``commonly taught
in textbooks'' conflates \emph{both} the early developers of quantum
theory \emph{and} the varied modern expositions of it into a vague,
undifferentiated mishmash.  Asher Peres' textbook is more
instrumentalist than the undergraduate standards~\cite{Peres-book};
the \emph{Feynman Lectures} handle probability in a less frequentist
way than Peres~\cite{FeynmanLP}.  Are all common textbooks Copenhagen, or is
Copenhagen that which is commonly taught in all textbooks?

The general disorganized sloppiness of the ``Quantum Bayesianism''
page illustrates what happens when articles grow by accretion: many
small additions, contributed by multiple authors without an overall
plan, forming a text by bricolage.  Volunteer collaboration can do
great things, but just as not all computational problems admit an easy
parallelization, not all writing tasks are well served by the
Wikipedia approach.

``Wander off the big ideas in the sciences,'' writes one observer,
``and you're likely to run into entries that are excessively technical
and provide almost no context, making them effectively
incomprehensible''~\cite{Timmer2015}.  Of course, technical content
has its place, but the material that would lead into the
technicalities, outlining the motivations and prerequisites for them,
isn't there.  It doesn't get written, because that kind of work is
\emph{harder to do} by piecemeal additions.  Wikipedia relies on
``drive-by contributors''~\cite{Shulman2016} working in their copious
free time with only the haziest of overall plans, and with essentially
no opportunity to get academic recognition for their effort.  This
puts a great barrier in the way of building pedagogically useful
material.

The question of which topics get covered, and to what extent and with
what level of sophistication, is a fascinating one~\cite{Giaimo2016}.
(At least, I have puzzled over it for years.)  But what happens when
the material that \emph{does} get written contains errors?

Obvious disruption on Wikipedia, like replacing entire articles with
strings of profanity, gets caught and reversed quickly.  Subtle
changes last longer and propagate farther.  Some silliness about the
Golden Ratio and Duchamp's \booktitle{Nu descendant un escalier
  n$^\circ$~\!2} survived for over a year, and when it was caught, the
``fix'' resulted in attributing a claim to a source that did not
contain it~\cite{sunclipse-duchamp}.  A couple of stoned college kids
made a joking edit to the page of a children's book author, which
endured for over five years and ended up uncritically accepted in,
among many other places, a book about Jesus~\cite{Dickson2014}.  A
purported Australian aboriginal deity, ``Jar'Edo Wens,'' lasted almost
ten years and appeared in a book promoting atheism, among a long list
of gods who had faded from belief~\cite{Dewey2015}.

Sometimes, the process even closes back upon itself: careless writers
elsewhere use a bit of entertaining trivia invented on Wikipedia, and
their writings in turn become sources to fill in the ``citation
needed'' tag on the Wikipedia page.  This is how the coati became the
``Brazilian aardvark''~\cite{Randall2014}.  Similarly, thanks to other
sites mirroring Wikipedia content, the Wikipedia page for ``Pareto
efficiency'' was self-referential for three months~\cite{Granade2014}.
The cartoonist Randall Munroe termed this process
``citogenesis''~\cite{Munroe2011}, and the results can be remarkably
difficult to sort out, particularly if the debunking requires
``Original Research.'' In the Wikipedian argot, ``OR'' includes ``any
analysis or synthesis of published material that serves to reach or
imply a conclusion not stated by the sources''~\cite{WPNOR}.  If no
source explicitly states that ``Jar'Edo Wens'' is a hoax or that the
``Brazilian aardvark'' is a prank, then arguing that they should be
removed easily spills over into ``OR.''

And so, Wikipedia's own policies, well-intentioned as they are, can
provide the amber to preserve fabrications both whimsical and
malicious.  For good or ill, the flies endure.

\begin{thebibliography}{999}
\bibitem{Lucretius} Lucretius, \booktitle{De Rerum Natura} (c.\ 60--50
  BCE), Book IV.  The relevant passage begins with, ``Invenies primis
  ab sensibus esse creatam / notitiem veri neque sensus posse
  refelli'' [You will find the knowledge of truth is created first by
    the senses, and the senses cannot be denied].
\end{thebibliography}
\end{document}